\RequirePackage{fix-cm}
\documentclass[twocolumn]{svjour3}          
\smartqed  
\usepackage{graphicx}
\usepackage{amsmath}
\usepackage{subfig}
\usepackage[colorlinks=true]{hyperref}
\begin{document}
\title{Two-breather solutions for the class I infinitely extended nonlinear Schr\"odinger equation and their special cases
}
\author{M. Crabb and N. Akhmediev}
\institute{M. Crabb \at
              Optical Sciences Group, Research School of Physics and Engineering, The Australian National University, Canberra, ACT, 2600, Australia.
              \email{matthew.crabb@anu.edu.au}}

\date{Received: date / Accepted: date}

\maketitle

\begin{abstract}
We derive the two-breather solution of the class I  infinitely extended nonlinear Schr\"odinger equation. We present a general form of this multi-parameter solution that includes infinitely many free parameters of the equation and free parameters of the two-breather components. Particular cases of this solution include rogue wave triplets, and special cases of 'breather-to-soliton' and `rogue wave-to-soliton' transformations. The presence of many parameters in the solution allows one to describe wave propagation problems with higher accuracy than with the use of the basic NLSE.
\end{abstract}

\section{Introduction}
\label{intro}
	
The  nonlinear Schr\"{o}dinger  equation \cite{Zakharov,bk} (NLSE) has various applications in describing ocean waves \cite{Osborne,KPS,Onorato}, pulses in optical fibres \cite{Hasegawa,agraw,Dudley}, Bose-Einstein condensates \cite{Konotop,Bobrov,Galati,Mithun}, waves in the atmosphere \cite{Stenflo}, plasma \cite{Kolomeisky} and many other physical systems \cite{Sulem,Yildirim,Castro,Czachor,Vladimirov}. Various extensions of the NLSE have been considered \cite{Roelofs,Ohkuma,Qurashi} that increase the accuracy of the description of nonlinear wave phenomena in these systems by incorporating higher-order effects \cite{Mihalache2,Mihalache3,Baronio,Andreev}. Higher-order terms in these extensions are responsible for linear dispersion, as well as nonlinear effects such as self-phase modulation, pulse self-steepening, the Raman effect, and so on \cite{agraw,Trippenbach}. These higher-order terms are important in nonlinear optics \cite{Potasek,Cavalcanti}, ocean wave dynamics \cite{Trulsen,Slunyaev,Sedlet} and especially in modelling high-amplitude rogue wave phenomena \cite{Onorato2,ds-oyI,Baronio2013}.

Adding higher-order terms generally results in the loss of integrability of the resulting equation. This means that exact solutions cannot be written in analytical form, making the treatment more complicated. However, a special choice of the higher-order operators in these extensions allows us to keep the integrability. The power of using these operators consists in the possibility of applying arbitrary real coefficients to each of these operators, thus significantly extending the range of physical problems that can be solved in exact form. It was found that the NLSE can be extended to arbitrarily high orders of these operators \cite{Chaos2015,PRE16,PRE17}, and these operators have been explicitly presented up to eighth order \cite{PRE16}. Using their recurrence relations, they can be calculated to any order, although the explicit form quickly becomes cumbersome. Nevertheless, there are no conceptual difficulties in construction of these equations. Moreover, infinitely many terms can be considered when finding solutions of these equations.

Presently, there are two sets of these operators that can be used for infinite-order extensions of the NLSE. 
We call them the class I \cite{Chaos2015,PRE16,PRE17} and class II \cite{CHAOS2018,Naturforsch2018} infinite extensions of the NLSE \cite{Crabb2019}. The presence of two independent extensions enables the more accurate description of physical problems with greater flexibility. Here, we deal exclusively with the class I extension. The class II extension is more involved, and will be left beyond the scope of the present work.\\\indent
In this paper, we find 2-breather solutions of the class I infinitely extended NLSE equation. These are multi-parameter solutions that involve both the free parameters of the equation, and free parameters of the solution, which together control the features of the two breather components, such as their localisation, propagation, and their relative position and frequencies. 
The presence of an infinite number of free parameters allows us to consider many particular cases, such as breather-to-soliton conversion, which is exclusive to higher-order extensions of the basic equation.

We also derive several limiting cases, the most important one of which is the general second-order rogue wave solution, a particular case of the 2-breather collision. However, only a limited number of special cases can be given in the frame of a single manuscript. We leave others for future work in this direction.

\section{The class I infinitely extended NLSE}
First, we give a brief exposition of the class I infinitely extended nonlinear Schr\"odinger equation. It is the integrable equation written in general form \cite{Chaos2015,PRE16}
\begin{equation}
\label{orig}
i\psi_x+F(\psi,\psi^*)=0,
\end{equation} 
where the operator $F(\psi,\psi^*)$ is defined through
\begin{equation}\label{orig2}
F=\sum_{n=1}^{\infty}(\alpha_{2n}K_{2n}-i\alpha_{2n+1}K_{2n+1}),
\end{equation} 
with the operators $K_n$ defined recursively by the integrals of the nonlinear Schr\"odinger equation \cite{Chaos2015}, and where each coefficient $\alpha_n$ is an arbitrary real number; that is,
\[K_n(\psi,\psi^*)=(-1)^n\frac{\delta}{\delta\psi^*}\int p_{n+1}dt,\]
where $p_n$ is the $n$-th integral of the basic nonlinear Schr\"odinger equation, and $p_{n+1}$ can be defined recursively as
\[p_{n+1}=\psi\frac{\partial}{\partial t}\left(\frac{p_n}{\psi}\right)+\sum_{r=1}^{n}p_{n-r}p_r,~p_1=|\psi|^2.\]
The four lowest order operators $K_n$ $(n=2,3,4,5)$ derived in this way are:
\begin{eqnarray}
\nonumber
K_2(\psi,\psi^*)&=& \psi_{tt}+2|\psi|^2\psi   ,
\\  
\nonumber
K_3(\psi,\psi^*)&=& \psi_{ttt}+6|\psi|^2\psi_t   ,
\\  \nonumber
K_4(\psi,\psi^*)&=&  \psi _{tttt}+8|\psi|^2\psi_{tt}+6|\psi|^4\psi+
\\  \nonumber
&& +4\psi|\psi_t|^2+6\psi_t^2\psi^*+2\psi^2\psi^*_{tt}. 
\\   \nonumber
K_5(\psi,\psi^*) &=&   \psi _{ttttt} +10|\psi|^2  \psi_{ttt}+10(\psi\,|\psi_{t}|^2)_t+
\\ \label{mkdv}
&& +   20 \psi^*  \psi_t \psi _{tt}+30|\psi|^4\psi_t.
\end{eqnarray}  
A few others can be found in \cite{PRE17}.
The operators $K_n$ involve linear terms with derivatives of order $n,$ and nonlinear terms involving $t$-derivatives of the function $\psi$ and its complex conjugate $\psi^*$. \\\indent
As already mentioned, the numbers $\alpha_n$ can take any values whatsoever, and do not need to be viewed as representing small perturbations for the equation (\ref{orig}) to be completely integrable.  This allows us to find solutions for which any order of dispersion can be taken into account without the need for approximation or numerical techniques. This extension substantially widens the range of applicability of the NLSE for solving nonlinear wave evolution problems.\\\indent
When only $\alpha_2\neq0,$ we have the fundamental, or `basic' nonlinear Schr\"odinger equation:
\begin{equation}
i\psi_x+\alpha_{2}K_2(\psi,\psi^*)=i\psi_x+\alpha_{2}(\psi_{tt}+2|\psi|^2\psi)=0,
\end{equation}
which includes the lowest-order dispersion and self-phase modulation terms. Further, if only $\alpha_2$ and $\alpha_3$ are nonzero, we have the integrable Hirota equation \cite{hirota}:
\begin{equation}\label{hir} 
i\psi_x+\alpha_{2}(\psi_{tt}+2|\psi|^2\psi)-i\alpha_3(\psi_{ttt}+6|\psi|^2\psi_t)=0.
\end{equation}	
Adding the fourth-order operator, $K_4,$ into  Eq.(\ref{hir}), gives the Lakshmanan-Porsezian-Daniel (LPD) equation \cite{Por88,porsezian1992}, and so on.

Again, the coefficients $\alpha_n$ are finite and arbitrary. However, physical applications, in general, require dispersive effects to decrease rapidly in strength with increasing order $n$. Convergence will thus not be an issue in practice for series involving $\alpha_n,$ and we will therefore be comfortable leaving the operator $F$ for the whole equation (\ref{orig}), as well as any other associated parameters, in the form of an infinite series when necessary.
\\\indent
While the operators $K_n$ in (\ref{mkdv}) rapidly become more complicated and the resulting differential equation of order $n$ becomes much harder to solve, exact solutions can be found explicitly by using already known solutions to the NLSE as a guide, and a large class of breather and soliton solutions are already known \cite{PRE16,PRE17}. In previous works \cite{PRE16,PRE17}, we have seen that the effect of nonzero odd order operators is to transform $t$ as $t\mapsto t+vx$ with $v$ being a function of all coefficients $\alpha_{2n+1}$. The effect of the nonzero even order operators is to transform $x$ as $x\mapsto Bx,$ with $B$ being a function of the parameters $\alpha_{2n}$.

In this work we extend this approach to a general family of second-order solutions, so we introduce parameters $B_1$ and $B_2,$ and $v_1$ and $v_2$, to play an analogous role for the two distinct breather components. This enables us to generalise the 2-breather to the infinite extension of the NLSE, and we now proceed to the analysis of these solutions.


\section{The 2-breather solution}
\label{sec:1}
\begin{figure}[ht]
	\includegraphics[scale=0.32]{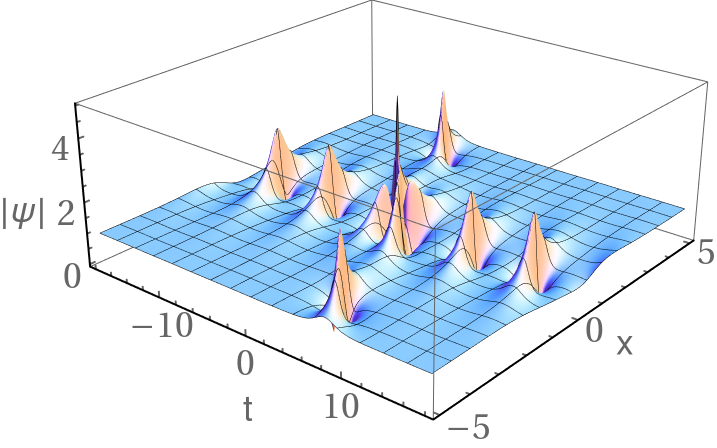}
	\caption{\it{A collision between an Akhmediev breather and Kuznetsov-Ma soliton, with $\kappa_1=1,$ and $\kappa_2=i.$ Here $\alpha_n=1/n!$ up to $n=8,$ with all higher $\alpha_n=0.$ 
	}}
	\label{2BE}
\end{figure}

\begin{figure}[ht]
	\includegraphics[scale=0.32]{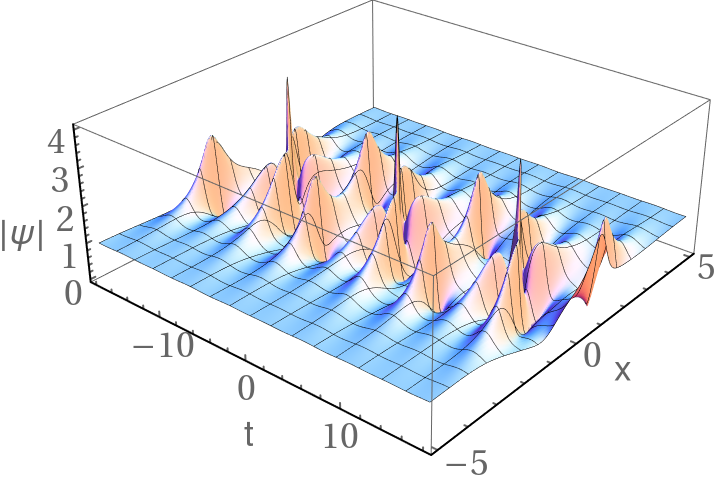}
	\caption{\it{The 2-breather solution with $\alpha_n=1/n!$ up to $n=8,$ with all higher $\alpha_n=0,$ but now with $\kappa_1=\tfrac32,$ and $\kappa_2=1.$}}
	\label{32B}
\end{figure}

Higher analogues of the Akhmediev breathers can be obtained through iterations of the Darboux transformation \cite{2-sol,kedb}. After transforming the plane wave solution $e^{ix}$ with a Darboux transformation, with an eigenvalue $\lambda$ such that $\lambda^2\neq-1,$ and repeating this transformation twice, we get the 2-breather solution to the basic NLSE. This can then be generalised to the 2-breather solution of the extended equation. The general 2-breather solution is of the form
\begin{equation}\label{basicsol}
\psi(x,t)=\left\{1+\frac{G(x,t)+iH(x,t)}{D(x,t)}\right\}e^{i\phi x},
\end{equation}
where
\begin{eqnarray}
G(x,t)&=&-(\kappa_1^{\;2}-\kappa_2^{\;2})\bigg\{\frac{\kappa_1^{\;2}\delta_2}{\kappa_2}\cosh\delta_1B_1x\cos\kappa_2t_2-\nonumber\\
&-&\frac{\kappa_2^{\;2}\delta_1}{\kappa_1}\cosh\delta_2B_2x\cos\kappa_1t_1-\nonumber\\
&-&(\kappa_1^{\;2}-\kappa_2^{\;2})\cosh\delta_1B_1x\cosh\delta_2B_2x\bigg\},\\
H(x,t)&=&-2(\kappa_1^{\;2}-\kappa_2^{\;2})\bigg\{\frac{\delta_1\delta_2}{\kappa_2}\sinh\delta_1B_1x\cos\kappa_2t_2-\nonumber\\
&-&\frac{\delta_1\delta_2}{\kappa_1}\sinh\delta_2B_2x\cos\kappa_1t_1-\nonumber\\
&-&\delta_1\sinh\delta_1B_1x\cosh\delta_2B_2x+\nonumber\\
&+&\delta_2\cosh\delta_1B_1x\sinh\delta_2B_2x\bigg\},\\
D(x,t)&=&2(\kappa_1^{\;2}+\kappa_2^{\;2})\frac{\delta_1\delta_2}{\kappa_1\kappa_2}\cos\kappa_1t_1\cos\kappa_2t_2 +4\delta_1\delta_2\times  \nonumber \\
&\times& (\sinh\delta_1B_1x\sinh\delta_2B_2x+\sin\kappa_1t_1\sin\kappa_2t_2)-\nonumber\\
&-&(2\kappa_1^{\;2}-\kappa_1^{\;2}\kappa_2^{\;2}+\kappa_2^{\;2})\cosh\delta_1B_1x\cosh\delta_2B_2x-\nonumber\\
&-&2(\kappa_1^{\;2}-\kappa_2^{\;2})\bigg\{\frac{\delta_1}{\kappa_1}\cosh\delta_2B_2x\cos\kappa_1t_1-\nonumber\\
&-& \frac{\delta_2}{\kappa_2}\cosh\delta_1B_1x\cos\kappa_2t_2\bigg\},
\end{eqnarray}
Here $\kappa_1$ and $\kappa_2$ are the modulation parameters, 
\begin{equation*}
\delta_m=\tfrac12\kappa_m\sqrt{4-\kappa_m^{\;2}}
\end{equation*}
is the growth rate of the modulational instability for each breather component, and the shorthand notation $t_m$ indicates $t_m=t+v_mx$ for $m=1,2.$ Note that whenever $t_m$ appears, we have ignored a constant of integration, and we have also done the same whenever $\delta_mB_mx$ appears. The most general solution allows for the replacements $t_m\mapsto t_m-T_m$, and $\delta_mB_mx\mapsto\delta_mB_m(x-X_m),$ where $T_m$ and $X_m$ are real constants which determine relative positions along the axes of $t$ and $x,$ respectively, which we might include to incorporate a time delay in one breather component, for instance. For the time being, we set these constants to be both zero without substantial loss, to address their significance later.\\\indent
The phase factor $\phi$ is independent of the modulation, since this part has no physical effect on the modulation when it is real, and here it takes the same real value as it does for the plane wave solution, i.e.
\begin{equation}
\phi=\sum_{n=1}^{\infty}\binom{2n}{n}\alpha_{2n}.
\end{equation}
The values $B_m$ determine the modulation frequency of each component, and the parameters $v_m,$ although they cannot be considered velocities in the usual sense, introduce a tilt to $|\psi|$ relative to the axes of $x$ and $t.$ They are given explicitly by
\begin{align}
B_m&=\sum_{n=1}^{\infty}\binom{2n}{n}nF(1-n,1;\tfrac32;\tfrac14\kappa_m^{\;2})\alpha_{2n},\\
v_m&=\sum_{n=1}^{\infty}\binom{2n}{n}(2n+1)F(-n,1;\tfrac32;\tfrac14\kappa_m^{\;2})\alpha_{2n+1},
\end{align}
with $m=1,2,$ where $F(a,b;c;z)$ is the Gaussian hypergeometric function. Note that there is a simple relationship between $v_m$ and $B_m$: the coefficient of $\alpha_{2n}$ in $B_m$ is twice the coefficient of $\alpha_{2n-1}$ in $v_m.$
The first two terms of $B_m$ for the two-breather solution have been previously given in \cite{chowd}. Our new solution extends these coefficients to arbitrary orders of dispersion and nonlinearity.\\\indent
Also notice that the parameters are the same functions of $\kappa_m,$ for both $m=1,2.$ This is at least suggested by symmetry. If two successive Darboux transformations generate a 2-breather solution, then there must be two independent eigenvalues, corresponding to two independent modulation parameters. Physically, we could reason that there should be no way of knowing which breather component is which, so the order in which each component was generated by Darboux transformation should be equally irrelevant. If so, it should then follow that $B_1$ is the same function of $\kappa_1$ as $B_2$ is of $\kappa_2,$ and similar for $v_1$ and $v_2$, and this would also imply that $B_m$ and $v_m$ are the same functions as for the single-breather solution, which are already known \cite{PRE17}.
  It is worth considering whether this property extends to the general $n$-breather solution: i.e. whether, in general, we can find $B_1,\dots,B_n$ and $v_1,\dots,v_n$ which are the same functions of their respective modulation parameters $\kappa_1,\dots,\kappa_n,$ but we do not answer this question here.\\\indent
The growth rate $\delta_m$ in both components will be real when $\kappa_m$ is real, but the eigenvalues of the Darboux transformation are free to take any complex value at all, although the transformations are trivial when the eigenvalues are real, and thus so are the modulation parameters. Real-valued modulation parameters correspond to Akhmediev breathers, whereas imaginary-valued modulation parameters correspond to Kuznetsov-Ma solitons, the functional form of the breathers being otherwise equivalent. An example which shows the difference between real and imaginary modulation parameters is given in Fig. (\ref{2BE}). In Fig. (\ref{32B}) we give an example of the effects of altering the ratio of the modulation of the two components.

\begin{figure}[ht]
	\includegraphics[scale=0.32]{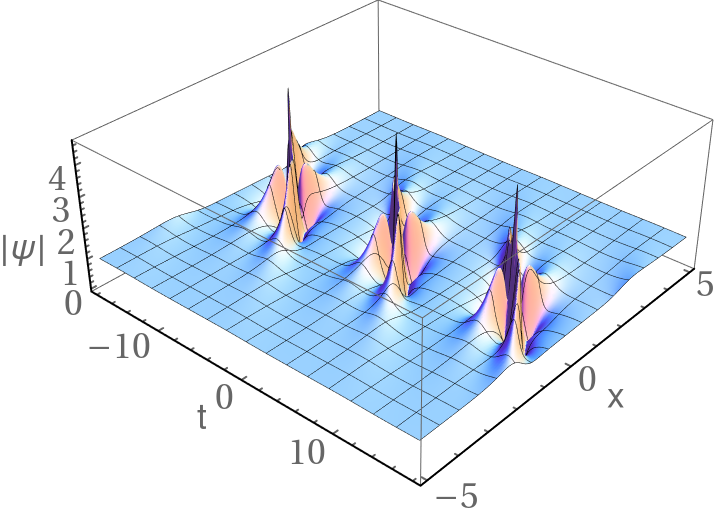}
	\caption{\it{The 2-breather solution (\ref{basicsol}) of Eq.(\ref{orig}). The modulation parameters are at a ratio $\kappa_1:\kappa_2=1:2$. Parameters of the equation are: $\alpha_2=\tfrac12,$ $\alpha_3=\tfrac16,$ $\alpha_4=\tfrac{1}{24},$ $\alpha_5=\tfrac{1}{30},$ $\alpha_6=\tfrac{1}{144},$ with all higher $\alpha_n=0.$ The wave profile is tilted in the $(x,t)$-plane due to the nonzero $v_m$.}}
	\label{2b}
\end{figure}

\section{Breather-to-soliton conversion}
\label{sec:2}
\begin{figure}[ht]
	\includegraphics[scale=0.32]{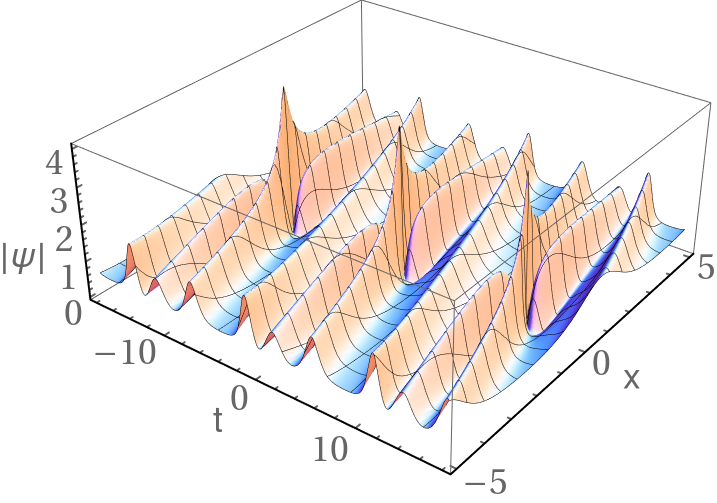}
	\caption{\it{A wave profile of a `breather-to-soliton conversion'. We use the same set of parameters as in Fig. \ref{32B}, except $\alpha_2$ is now chosen such that $B_2=0.$ This choice extends to infinity the tails of the breathers that would otherwise decay.}}
	\label{bsc}
\end{figure}

If we choose parameters $\alpha_n$ such that $B_m=0,$ the 2-breather solution may then behave in a way which is unique to the extension of the nonlinear Schr\"odinger equation \cite{chowd}, in the sense that it is only when higher orders of dispersion and nonlinearity are accounted for that it is possible to take $B_m=0$ without obtaining a trivial or otherwise degenerate solution.\\\indent
For example, if we choose $\alpha_2$ such that $B_2=0$ for all $\kappa_2,$ then writing $B_1=B,$ it is easy to show that $B$ must take the value
\[B=\sum_{n=1}^{\infty}\binom{2n+2}{n+1}(n+1)E \left(-n,1;\tfrac32;\tfrac14\kappa_1^{\;2}|\tfrac14\kappa_2^{\;2} \right)\alpha_{2n+2},\]
where we define the function $E$ as the difference of hypergeometric functions:
\[E(a,b;c;z_1|z_2)=F(a,b;c;z_1)-F(a,b;c;z_2).\]
\indent We can then simplify the general 2-breather solution considerably. We obtain
\begin{align}
G(x,t)&=(\kappa_1^{\;2}-\kappa_2^{\;2})\bigg\{\frac{\kappa_1^{\;2}\delta_2}{\kappa_2}\cosh\delta_1Bx\cos\kappa_2t_2-\nonumber\\
	&~~~-\frac{\kappa_2^{\;2}\delta_1}{\kappa_1}\cos\kappa_1t_1-(\kappa_1^{\;2}-\kappa_2^{\;2})\cosh\delta_1Bx\bigg\}\nonumber,
\end{align}
\begin{align}
H(x,t)&=2\delta_1(\kappa_1^{\;2}-\kappa_2^{\;2})\sinh\delta_1Bx\bigg(1-\frac{\delta_2}{\kappa_2}\cos\kappa_2t_2\bigg)\nonumber,
\end{align}
\begin{align}
D(x,t)&=2(\kappa_1^{\;2}+\kappa_2^{\;2})\frac{\delta_1\delta_2}{\kappa_1\kappa_2}\cos\kappa_1t_1\cos\kappa_2t_2+\nonumber\\
&~~~+4\delta_1\delta_2\sin\kappa_1t_1\sin\kappa_2t_2-\nonumber\\
&~~~-(2\kappa_1^{\;2}-\kappa_1^{\;2}\kappa_2^{\;2}+\kappa_2^{\;2})\cosh\delta_1Bx-\nonumber\\
&~~~-2(\kappa_1^{\;2}-\kappa_2^{\;2})\bigg(\frac{\delta_1}{\kappa_1}\cos\kappa_1t_1-\nonumber
\end{align}
\begin{align}
&~~~~~-\frac{\delta_2}{\kappa_2}\cosh\delta_1Bx\cos\kappa_2t_2\bigg).
\end{align} 
An example of this solution is given in Fig. \ref{bsc}. The difference of this solution from the one shown in Fig. \ref{2b} is that the wave profiles at $x\to\pm\infty$ are not plane waves. The periodic set of tails from each breather maximum extends to infinity, reminiscent of periodically repeating solitons. This is the phenomenon that is known as breather-to-soliton conversion \cite{chowd}. Clearly, these `solitons' do not have a separate spectral parameter related to them.
\section{The 2-breather solution in the semirational limit}
\label{sec:3}
\begin{figure}[ht]
	\includegraphics[scale=0.32]{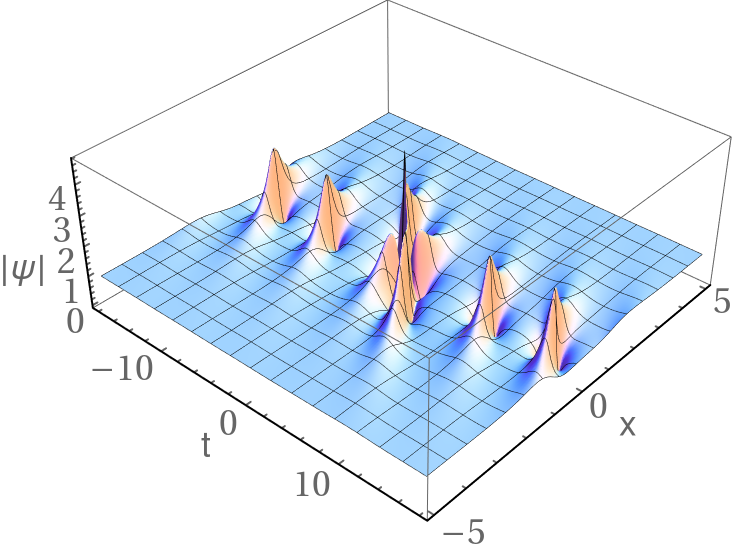}
	\caption{\it{The 2-breather solution in the semirational limit. Here the nonzero modulation parameter is $\kappa=1,$ with $\alpha_n$ the same as in Fig. (\ref{2b}). It can be considered as a superposition of the Akhmediev breather with the Peregrine solution.}}
	\label{2Bsr}
\end{figure}

When one of the modulation parameters, say $\kappa_2,$ tends to zero, we obtain the semirational limit, i.e. a solution obtained as a combination of polynomials and circular or hyperbolic functions. Then, writing $\kappa$ for $\kappa_1$, and $\delta$ for $\delta_1,$ the functions $G$, $H$, and $D$ become
\begin{eqnarray}
G(x,t)&=&\tfrac18\kappa^2\{\kappa^2(1+4t_2^{\;2}+4B_2^{\;2}x^2)-1\}\cosh\delta B_1x+ \nonumber\\
&+&\kappa\delta\cos\kappa t_1,\nonumber
\end{eqnarray}
\begin{eqnarray}
H(x,t)&=&2\kappa B_2x(\delta\cos\kappa t_1-\kappa\cosh\delta B_1x)+\nonumber\\
&+&\tfrac14\delta\kappa^2(1+4t_2^{\;2}+4B_2^{\;2}x^2)\sinh\delta B_1x\nonumber
\end{eqnarray}
\begin{eqnarray}
D(x,t)&=&\frac{\delta}{\kappa}\{4-\tfrac14\kappa^2(1+4t_2^{\;2}+4B_2^{\;2}x^2)\}\cos\kappa t_1+\nonumber\\
&+&4\delta B_2x\sinh\delta B_1x+\delta t_2\sin\kappa t_1-\nonumber\\
&-&\{4+\tfrac14\kappa^2(1+4t_2^{\;2}+4B_2^{\;2}x^2)\}\cosh\delta B_1x,
\end{eqnarray}
and the parameters $B_2$ and $v_2$ are reduced to
\begin{equation}
B_2=\sum_{n=1}^{\infty}\binom{2n}{n}n\alpha_{2n},
\end{equation}
and
\begin{equation}
v_2=\sum_{n=1}^{\infty}\binom{2n}{n}(2n+1)\alpha_{2n+1}.
\end{equation}
This semirational 2-breather solution is a superposition of a Peregrine solution with the Akhmediev breather, since taking the limit $\kappa_2\to0$ reduces the frequency of one of the breathers to zero, meaning that it is transformed to a Peregrine solution. A plot of this solution is shown in Fig. \ref{2Bsr}. Here, the central feature is roughly the second order rogue wave while the peaks away from the origin belong to the remaining first-order breather.

\section{The degenerate two-breather limit}
\label{sec:4}
If both eigenvalues of the Darboux transformation are taken to be equal, so that both modulation parameters $\kappa_m$ are also equal, we obtain the case of degenerate breathers. Direct calculations provide no solution. In this case, one modulation parameter should instead be taken as a small perturbation from the other, say $|\kappa_1-\kappa_2|=\varepsilon.$ Then, we take the limit as the perturbation $\varepsilon$ becomes arbitrarily small, so that the solution remains well-defined at all times. Namely, if we put $\kappa_1=\kappa,$ and $\kappa_2=\kappa+\varepsilon,$ we have
\[B_2=\sum_{n=1}^{\infty}\binom{2n}{n}nF(1-n,1;\tfrac32;\tfrac14(\kappa+\varepsilon)^2)\alpha_{2n},\]
and
\[v_2=\sum_{n=1}^{\infty}\binom{2n}{n}(2n+1)F(-n,1;\tfrac32;\tfrac14(\kappa+\varepsilon)^2)\alpha_{2n+1}.\]

\begin{figure}[ht]
	\includegraphics[scale=0.35]{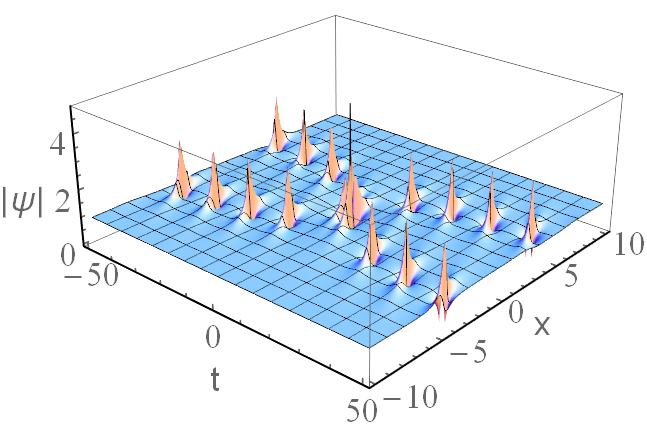}
	\caption{\textit{The degenerate 2-breather solution. We take the set of $\alpha_n$ the same as in Fig. \ref{2b}, and the modulation parameters $\kappa_1=\kappa_2=\tfrac12.$ The two breathers collide with the high peak at the origin due to the synchronised phases.}}
	\label{db1}
\end{figure}

Next, recalling the identity
\[\frac{d}{dz}F(a,b;c;z)=\frac{ab}{c}F(a+1,b+1;c+1;z),\]
take the Maclaurin series of the $G(x,t)$, $H(x,t)$, and $D(x,t)$ with respect to $\varepsilon.$ In the limit as $\varepsilon\to0,$ the ratio of these series will be a well-defined solution with equal eigenvalues; it is thus sufficient to consider only the lowest-order non-vanishing terms in the series expansion for $D(x,t)$ in $\varepsilon,$ which in this case happen to be the coefficients of $\varepsilon^2$. By this method we obtain the degenerate 2-breather solution in the form (\ref{basicsol}) with
\begin{eqnarray}
\label{degen}
G(x,t)&=&2\kappa^2\bigg[1+\cosh2\delta Bx+\bigg\{\bigg(\kappa B-\frac{2\delta^2}{\kappa}B-\nonumber\\
&-&\delta^2B'\bigg)x\sinh\delta Bx-\nonumber\\
&-&\frac{\kappa}{\delta}\left(1-\frac{\delta^2}{\kappa^2}\right)\cosh\delta Bx\bigg\}\cos\kappa(t+vx)-\nonumber\\
&-&\{t+(v+\kappa v')x\}\delta\cosh\delta Bx\sin\kappa(t+vx)\bigg]\nonumber,
\end{eqnarray}
\begin{eqnarray}
H(x,t)&=&2\kappa\bigg[\left\{\left(\frac{2\delta^2}{\kappa^2}-1\right)\kappa B+2\delta^2B'\right\}x+\nonumber\\
&+&\tfrac12\delta\bigg\{\tfrac12\left(\frac{2\delta^2}{\kappa^2}-1\right)Bx-\frac{\delta^2}{\kappa}B'\bigg\}x\cosh\delta Bx\times\nonumber\\
&\times&\cos\kappa(t+vx)+\frac{\kappa}{\delta}\bigg(\frac{2\delta^2}{\kappa^2}-1\bigg)\sinh2\delta Bx-\nonumber\\
&-&\delta^2\sinh\delta Bx\{\cos\kappa(t+vx)+\nonumber\\
&+&\kappa\sin\kappa(t+vx)\} \{t+(v+\kappa v')x\}\bigg]\nonumber,
\end{eqnarray}
\begin{eqnarray}
D(x,t)&=&\frac{\kappa^2}{32\delta^2}\bigg[-8\kappa^2\left(1+\frac{\delta^2}{\kappa^2}\right)-\frac{64\delta^4}{\kappa^2}(t+vx)^2-\nonumber\\
&-&64\delta^2\bigg(1-\frac{2\delta^2}{\kappa^2}\bigg)^2B^2x^2-32\cosh2\delta Bx-\nonumber\\
&-&\frac{128\delta^2}{\kappa}\bigg\{\bigg(2-\frac{4\delta^2}{\kappa^2}\bigg)B-\frac{\delta^2}{\kappa}B'\bigg\}x\sinh\delta Bx\times\nonumber\\
&\times&\cos\kappa(t+vx)-32\delta\bigg\{\kappa\cos\kappa(t+vx)+\nonumber \\
&+&\frac{4\delta^2}{\kappa^2}\{t+(v+\kappa v')x\}\sin\kappa(t+vx)\bigg\}\cosh\delta Bx+\nonumber 
\end{eqnarray}
\begin{eqnarray}
&+&\frac{16\delta^2}{\kappa^2}\bigg\{2\cos2\kappa(t+vx)+\bigg(8\bigg(1-\frac{2\delta^2}{\kappa^2}\bigg)\kappa BB'x-\nonumber\\
&-&4\delta^2B'^2\bigg)\kappa^2x-4v'\{2t+(v+\kappa v')x\}\bigg)\bigg\}\bigg],
\end{eqnarray}
where $B=B_1,$ and we use $B'$ and $v'$ to denote the partial derivatives of $B_2$ and $v_2$ with respect to $\varepsilon$ evaluated at the point $\varepsilon=0,$ i.e. when $\kappa_2\to\kappa.$ That is,
\[B'=\tfrac13\kappa\sum_{n=1}^{\infty}\binom{2n}{n}n(1-n)F(2-n,2;\tfrac52;\tfrac14\kappa^2)\alpha_{2n},\]
and
\[v'=-\tfrac13\kappa\sum_{n=1}^{\infty}\binom{2n}{n}(2n+1)nF(1-n,2;\tfrac52;\tfrac14\kappa^2)\alpha_{2n+1},\]
etc. We drop the subscripts due to the fact that as $\varepsilon\to0$ both modulation parameters take equal values anyway. 
A plot of this solution is given in Fig.\ref{db1}.

The degenerate breather solution is a one-parameter family of solutions which represents the collision of two breathers with the same modulation parameter $\kappa$, or, equivalently, with equal frequencies. It can be considered a generalisation of the known 2-soliton solution for the class I extension of the nonlinear Schr\"odinger equation \cite{zna}.

\section{Second-order rogue wave solution}
\label{sec:5}
\begin{figure}[ht]
	\includegraphics[scale=0.35]{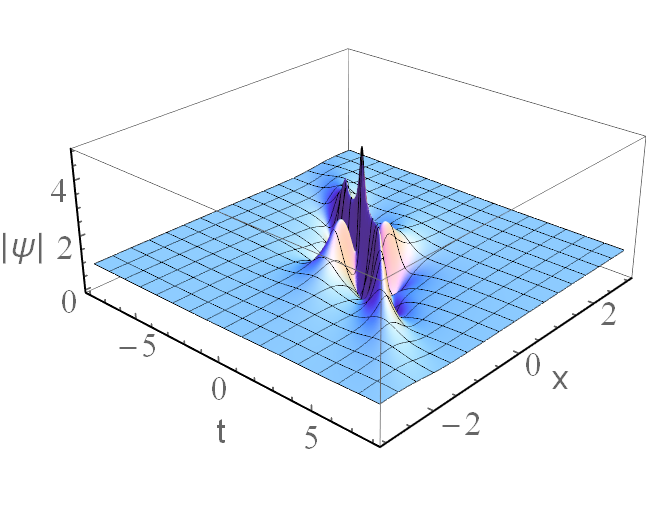}
	\caption{\textit{The second-order rogue wave, Eqs.(\ref{mrg}),(\ref{mrh}),(\ref{mrd}) obtained from the degenerate 2-breather solution shown in Fig. \ref{db1} in the limit $\kappa\to0,$ with some stretching due to higher-order effects.}}
	\label{r1}
\end{figure}

When the frequency of the degenerate breather tends to zero, the spacing between the successive peaks in Fig. \ref{db1} becomes infinitely large, pushing them out to infinity. What remains at the origin is the second-order rogue wave. In order to derive this solution, we take the limit $\kappa\to0$  in the expressions (\ref{degen}). However, calculations show that this limit cannot be found directly. In order to find it, we repeatedly apply l'H\^opital's rule to the degenerate breather solution as $\kappa\to0$. The derivatives of $G,$ $H,$ and $D$ with respect to $\kappa$ at the point $\kappa=0$ vanish up to $O(\varepsilon^6)$. The resulting functions $G$, $H,$ and $D$ become polynomials:
 \begin{align}
\label{mrg}
	G(x,t)&=12\{-3+24(3B^2-BB''-vv'')x^2+\nonumber\\
	&~~~+80B^4x^4-192v''xt+96B^2x^2(t+vx)^2+\nonumber\\
	&~~~+24(t+vx)^2+16(t+vx)^4\},\\
	\label{mrh}
	H(x,t)&=576B''x+2304B''x(t+vx)^2-24Bx\{15-\nonumber\\
	&~~~-8(B+16B'')Bx^2-16B^4x^4+\nonumber\\
	&~~~+192v''x(t+vx)-32B^2x^2(t+vx)^2+\nonumber\\
	&~~~+24(t+vx)^2-16(t+vx)^4\},\\
	\label{mrd}
	D(x,t)&=-9-36\{11B^2-48BB''+64B''^2-\nonumber\\
	&~~~-16(v-v'')v''\}x^2-48\{9B^4-6B^2v^2+\nonumber\\
	&~~~+16(3v^2-B^2)BB''+16(v^2-3B^2)vv''\}x^4-\nonumber\\
	&~~~-64(B^4+3B^2v^2+3v^4)B^2x^6+576v''xt-\nonumber\\
	&~~~-768v''xt^3+288(B^2-8BB''-8vv'')x^2t^2-\nonumber\\
	&~~~-192B^2x^2t^4+576\{(B-8B'')Bv+\nonumber\\
	&~~~+4(B^2-v^2)v''\}x^3t-768B^2vx^3t^3-\nonumber\\
	&~~~-192(B^2+6v^2)B^2x^4t^2-384(B^2+\nonumber\\
	&~~~+2v^2)B^2vx^5t-108(t+vx)^2-\nonumber\\
	&~~~-48(t+vx)^4-64(t+vx)^6,
\end{align}
where, in the same limit as $\kappa\to0,$
\begin{align*}
B&=\sum_{n=1}^{\infty}\binom{2n}{n}n\alpha_{2n},\\
v&=\sum_{n=1}^{\infty}\binom{2n}{n}(2n+1)\alpha_{2n+1}.
\end{align*}
The first-order derivatives $B'$ and $v'$ vanish as $\kappa\to0,$ but the second-order derivatives still remain, and in the limit as $\kappa\to0$ are
\begin{align*}
	B''&=-\frac13\sum_{n=1}^{\infty}\binom{2n}{n}n(n-1)\alpha_{2n},\\
	v''&=-\frac13\sum_{n=1}^{\infty}\binom{2n}{n}(2n+1)n\alpha_{2n+1}.
\end{align*}
 This solution is shown in Fig. \ref{r1}. It is, naturally, the second-order rogue wave, but slanted and rescaled in the $(x,t)$-plane relative to the second-order rogue wave of the NLSE \cite{Akhm1985,WANDT}.
\section{Rogue wave triplets}
\begin{figure}[ht]
	\includegraphics[scale=0.38]{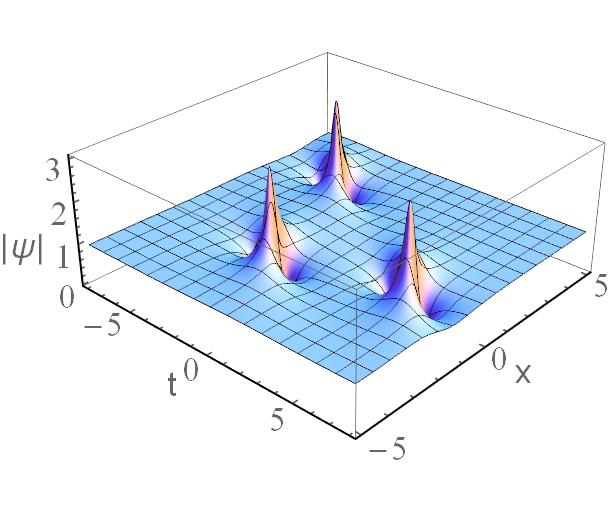}
	\caption{\textit{The second-order rogue wave triplet (\ref{trip}), with separation parameters $\xi=-\eta=480,$ and the extended equation parameters given by $\alpha_2=\tfrac12,$ $\alpha_3=\tfrac{1}{27},$ $\alpha_4=\tfrac{1}{50},$ $\alpha_5=\tfrac{1}{81},$ $\alpha_6=\tfrac{1}{200},$ $\alpha_7=\tfrac{1}{343}$. For this choice of parameters, we have $B=\tfrac{77}{50},$ $v=\tfrac{1324}{1323},$ $B''=-\tfrac{7}{25},$ and $v''=-\tfrac{2894}{3969}.$}}
	\label{trips}
\end{figure}

It is well known that the general $n$-th order rogue wave has the remarkable property of being able to split into $\tfrac12n(n+1)$ first-order components \cite{tri}. The second-order rogue wave discussed above is only a particular case of a more general rogue wave structure, where all three first-order components are located at the origin, and have merged into one single peak. In order to obtain the more general solution where the three components are not merged together, known as the rogue wave triplet \cite{triplet}, we re-introduce the constants of integration into the general 2-breather solution, i.e.
\begin{align*}
	\delta_mB_mx&\mapsto \delta_m(B_mx-\varepsilon X_m),\\
	t+v_mx&\mapsto t-T_m\varepsilon+v_mx,
\end{align*}
where $X_m$ and $T_m$ are arbitrary, and the parameter $\varepsilon$ is introduced to make sure that the Taylor series in the degenerate limit still vanishes up to $O(\varepsilon^2).$ The values of $X_m$ and $T_m$ determine the location of the components of the 
breather components. They add additional free parameters to the solution which we have previously given for the restricted case in which $X_m=T_m=0.$ Notice also that we do not make the replacement $x\mapsto x-X_m\varepsilon$ directly, but, for simplicity's sake, instead define $X_m$ to account for the higher-order terms in $B_m$.\\\indent
In order to further simplify parametrisation, we assume that $X_m$ and $T_m$ are functions of the modulation parameter $\kappa,$ and are of the order $O(\kappa)$. Then, defining free parameters $\xi$ and $\eta$ independent of $\kappa,$ such that 
\begin{align*}
	\kappa\xi&=48(X_1-X_2),\\
	\kappa\eta&=48(T_1-T_2),
\end{align*}
we have in the limit as $\kappa\to0$ the rogue wave triplet solution in the form
\begin{equation}
\label{trip}
	\psi(x,t)=\left\{1+\frac{\hat{G}(x,t)+i\hat{H}(x,t)}{\hat{D}(x,t)}\right\}e^{i\phi x},
\end{equation}
with
\begin{eqnarray}
\hat{G}(x,t)&=&G(x,t)-48\xi Bx-48\eta(t+vx),\\
\hat{H}(x,t)&=&H(x,t)+12\xi-48\xi B^2x^2-\nonumber\\
&-&96\eta Bx(t+vx)+48\xi(t+vx)^2,\\
\hat{D}(x,t)&=&D(x,t)-(\xi^2+\eta^2)+12\{\xi(3B-4B'')-\nonumber\\
&-&4\eta v''\}x+16\xi B^3x^3+12\eta(1+4B^2x^2)\times\nonumber\\
&\times&(t+vx)-48\xi Bx(t+vx)^2 \nonumber \\
&-&16\eta(t+vx)^3,
\end{eqnarray}
where $\hat{G},$ $\hat{H}$ and $\hat{D}$ now contain two new free parameters, $\xi$ and $\eta,$ which determine the separation of the fundamental rogue wave components in the triplet \cite{triplet}, and where $G,$ $H,$ and $D$ are as given in Eqs. (\ref{mrg})-(\ref{mrd}), for the particular case in which $\xi=\eta=0$.  An example of the formation of rogue wave triplets, corresponding to nonzero $\xi$ and $\eta,$ is shown in Fig. \ref{trips}. When both $\xi=0$ and $\eta=0$, all three components merge at the origin, as in Fig. \ref{r1}.

Generally, the coefficient $B$ in Eq. (\ref{trip}) determines the degree of localisation along the $x$-axis. Larger values of $B$ will correspond to narrower peaks, whereas smaller values of $B$ will correspond to broader peaks, and $B=0$ to minimal localisation. A point of interest here is that it is again possible to choose a parametrisation for which $B$ is any fixed constant. If we choose, for instance,
\[\alpha_2=c-\frac12\sum_{n=1}^{\infty}\binom{2n+2}{n+1}(n+1)\alpha_{2n+2},\]
we end up with $B=c,$ where $c$ is a free parameter. However, $B''$ is entirely independent of the choice of $c$, since the coefficient of $\alpha_2$ in $B''$ is zero. As the simplest example, we consider the completely de-localised case, $B=0,$ with $B''$ remaining arbitrary. The rogue wave solution then reduces to (\ref{trip}) with
\begin{align*}
\hat{G}(x,t)&=G_0(x,t)-48\eta(t+vx),\\
\hat{H}(x,t)&=H_0(x,t)+12\xi+48\xi(t+vx)^2,\\
\hat{D}(x,t)&=D_0(x,t)-(\xi^2+\eta^2)-48(\xi B''+\eta v'')x+\nonumber\\
&~~~+12\eta(t+vx)-16\eta(t+vx)^3.	
\end{align*}
where
\begin{align*}
G_0(x,t)&=12\{-3-192v''x(t+vx)+24(t+vx)^2+\nonumber\\
&~~~+16(t+vx)^4\},\\
H_0(x,t)&=576B''x\{1+4(t+vx)^2\},\\
D_0(x,t)&=-9-576\{4B''^2+v''^2\}x^2+576v''x(t+vx)-\nonumber\\
&~~~-768v''x(t+vx)^3-108(t+vx)^2-\nonumber\\
&~~~-48(t+vx)^4-64(t+vx)^6.
\end{align*}
Here, $G_0,$ $H_0,$ and $D_0$ are as given for the case where the components are merged and $B=0$, and $\hat{G},$ $\hat{H},$ $\hat{D}$ incorporate the shifting of the first-order components through $\xi$ and $\eta.$ 

\begin{figure}[htb]
	\includegraphics[scale=0.35]{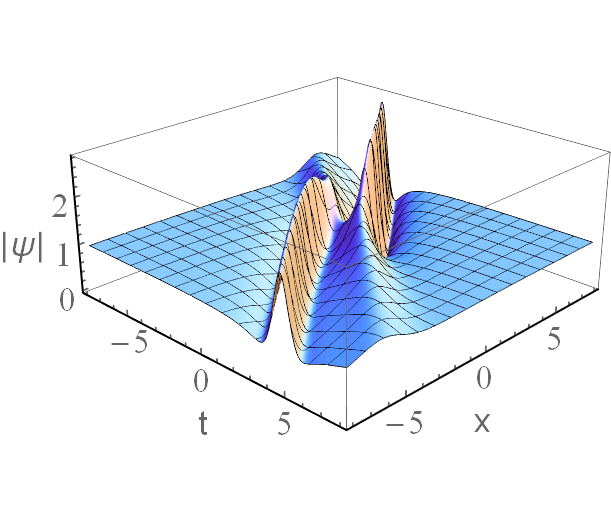}
	\caption{\textit{The second-order rogue wave solution  with `soliton'-like tails when $B=0,$ but now $\xi=\eta=48$.}}
	\label{zb1}
\end{figure}

When $B=0$, the second-order rogue wave acquires soliton-like tails similar to those in Fig. \ref{bsc}.  
\begin{figure}[htb]
	\includegraphics[scale=0.35]{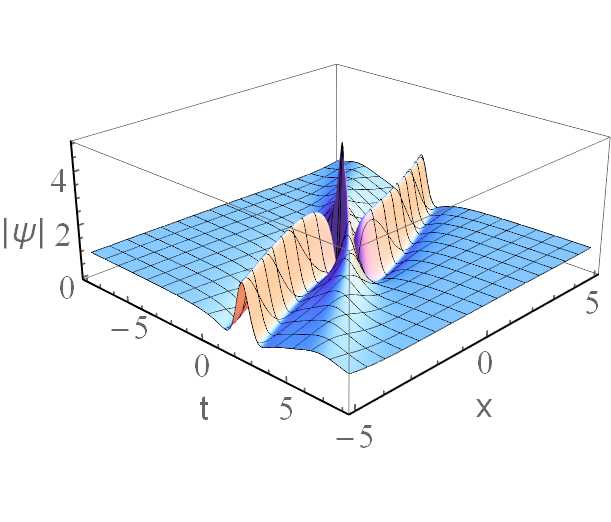}
	\caption{\textit{The second-order rogue wave solution with `soliton'-like tails when $\alpha_2$ chosen such that $B=0,$ and $\xi=\eta=0.$ Other parameters are the same as in Fig. \ref{trips}.}}
	\label{zb0}
\end{figure}
When, additionally, $\xi=\eta=0,$ rogue waves merge at the origin to form a second-order rogue wave with extended tails. This case is shown in Fig. \ref{zb0}. When $\xi$ or $\eta$ is not zero, the components split, resulting in the disappearance of the central peak. This case is shown in Fig. \ref{zb1}. Here, the central peak is absent but the long tails remain, consisting of de-localised first-order components.

\section*{Conclusions}
We have derived the general 2-breather solution for the class I infinitely extended nonlinear Schr\"odinger equation, and given many limiting cases; namely, breather-to-soliton conversions, the semirational limit, the degenerate 2-breather, and, probably most importantly, the general second-order rogue wave solution. These solutions completely describe a large family of second-order solutions to the class I extension of the NLSE, and exhibit rich behaviour.\\\indent
These results provide a more detailed analysis of the formation of nonlinear wave structures such as breathers and rogue waves when higher-order effects come into play, and leaves a large range of future related work wide open.
\begin{acknowledgements}
The authors gratefully acknowledge the support of the Australian 
Research Council (Discovery Project DP150102057).
\end{acknowledgements}

\end{document}